\documentclass[conference,a4paper]{APSIPA2025}
\usepackage{amsmath}
\usepackage{graphicx}
\usepackage{multirow}
\usepackage{threeparttable}

\usepackage{geometry}
\usepackage{comment}
\usepackage{caption}
\usepackage{subcaption}
\usepackage{balance}
\usepackage{multirow}
\usepackage{nicefrac, xfrac}

\usepackage{fancyhdr}
\usepackage{amssymb}

\fancypagestyle{firststyle}{
  \fancyhf{}
  \fancyhead[C]{2025 Asia Pacific Signal and Information Processing Association Annual Summit and Conference (APSIPA ASC)}
}

\begin{document}

\title{Parameter-Efficient Fine-Tuning of Foundation Models for CLP Speech Classification}

\author{
\authorblockN{
Susmita Bhattacharjee\authorrefmark{1}, 
Jagabandhu Mishra\authorrefmark{2},
H.S. Shekhawat\authorrefmark{1} and
S. R. Mahadeva Prasanna\authorrefmark{3}
}

\authorblockA{
\authorrefmark{1}
Indian Institute of Technology Guwahati, Guwahati \\
E-mail: sbhattacharjee@iitg.ac.in}

\authorblockA{
\authorrefmark{2}
University of Eastern Finland, Joensuu \\
}


\authorblockA{
\authorrefmark{3}
Indian Institute of Information Technology Dharwad, Dharwad \\
}
}

\maketitle
\thispagestyle{firststyle}
\pagestyle{fancy}

\begin{abstract}
We propose the use of parameter-efficient fine-tuning (PEFT) of foundation models for cleft lip and palate (CLP) detection and severity classification. In CLP, nasalization increases with severity due to the abnormal passage between the oral and nasal tracts; this causes oral stops to be replaced by glottal stops and alters formant trajectories and vowel space. Since foundation models are trained for grapheme prediction or long-term quantized representation prediction, they might have the potential to better discriminate CLP severity when fine-tuned on domain-specific data. We conduct experiments on two datasets: English (NMCPC) and Kannada (AIISH). We perform a comparative analysis using embeddings from self-supervised models Wav2Vec2 and WavLM, and the weakly supervised Whisper, each paired with SVM classifiers, and compare their results with traditional handcrafted features—eGeMAPS and ComParE. Finally, we fine-tune the best-performing Whisper model using PEFT techniques: Low-Rank Adapter (LoRA) and Decomposed Low-Rank Adapter (DoRA). Our results demonstrate that the proposed approach for severity classification achieves relative improvements of $26.4\%$ and $63.4\%$ in macro-average F1 score over the best foundation model and handcrafted feature baselines on the NMCPC dataset, and improvements of $6.1\%$ and $52.9\%$ on the AIISH dataset, respectively.

\end{abstract}

\section{Introduction}

Cleft lip and palate (CLP) is a congenital condition affecting the craniofacial region~\cite{evaluationand, methofperceptualassessment, clpnatureandremediation}. By $2019$, a total of $192,708$ cases of CLP had been reported globally~\cite{global}. CLP alters the structure of the oral and nasal cavities, which profoundly impacts speech production~\cite{evaluationand, methofperceptualassessment, clpnatureandremediation}. In particular, the opening between the lip and nasal cavity results in speech that is breathy and highly nasalized~\cite{Zajac2011ReliabilityAV, Whitehill2004SinglewordII}.  In addition, the opening between the oral and nasal cavities shifts the natural resonances of speech~\cite{evaluationand, kummer2007}, making it acoustically distinct from that of normal speakers. To treat this speech disorder effectively, the severity of CLP needs to be accurately detected and graded, which continues to be an active area of research in the medical domain~\cite{sensitivity}.

Researchers aim to detect and grade the severity of CLP directly from speech signals, motivated by the fact that CLP is a speech disorder~\cite{dataaugpro}. In individuals with CLP, the opening between the nasal and oral cavities primarily disrupts articulation and may also alter excitation characteristics~\cite{{dhananjaya2008speaker,sarma2015speaker}}. Compared to normal speech, CLP speech often exhibits pressure leakage from the oral to nasal tract during articulation, which leads speakers to replace oral stops such as /p/, /t/, and /k/ with glottal stops and to produce nasalized vowels~\cite{kummer2007}. Speakers with CLP also tend to produce distorted consonants, weak pressure sounds, and imbalanced resonance~\cite{clpbook}. These speech patterns introduce measurable changes in acoustic features, including increased nasal formants, altered formant trajectories, reduced vowel space, and distinctive spectral patterns associated with glottal or pharyngeal substitutions~\cite{dhananjaya2008speaker, kummer2007}.

Although research in this area remains limited, several studies have actively explored the automatic detection and severity classification of CLP speech~\cite{posterior, sireesha, 3cls, 3cls2, hypernasality, dcnn, Mathad2020ADL, jawed}. Researchers typically follow two main approaches: one applies \emph{signal processing} techniques guided by the articulatory and acoustic characteristics of CLP speech, while the other uses data-driven methods such as \emph{self-supervised or weakly supervised} foundation models to extract high-level representations directly from raw audio for classification tasks.~\cite{sireesha} extracted Mel-frequency cepstral coefficients (MFCCs) from variational mode decomposed signals to assess CLP severity. In \cite{3cls}, the authors parameterized the linear prediction residual signal on the vowel /i/ and extracted features such as vocal tract constriction, peak-to-sidelobe ratio, and spectral moments. The study in~\cite{dcnn} trained a deep neural network to generate aggregated posteriors of nasalized phonemes. Other works used automatic speech recognition posteriors~\cite{posterior}, representations from wav2vec2 models~\cite{germany}, and applied vision transformers~\cite{Nantha2025EnhancedCL}  to detect and assess the severity of CLP speech. In summary, data-driven methods and methods that utilize representations from foundation models have shown more promising performance compared to traditional signal processing based approaches~\cite{germany}.

Motivated by the promising performance of data-driven approaches, we present a comprehensive analysis of how representations from foundation models specifically, self-supervised (Wav2Vec2~\cite{wav2vec2}, WavLM~\cite{wavlm}), and weakly supervised (Whisper~\cite{whispersmall}) perform in detecting and classifying the severity of CLP speech. While researchers have conducted similar studies in other speech-based biomedical domains such as dementia~\cite{dementia1} and dysarthria~\cite{dysar} assessment, no such analysis, to the best of our knowledge, exists for CLP speech. Foundation models are typically trained to predict either \emph{quantized speech units} over long-term dynamics or \emph{speech graphemes} from normal, unimpaired speech. As a result, their learned representations may inherently carry discriminatory features that can distinguish CLP speech from normal speech and capture its severity. However, since most foundation models are trained on adult speech~\cite{wav2vec2, whispersmall, wavlm}, and CLP speech data predominantly comes from children~\cite{kummer2007}, this age-domain mismatch, along with the limited availability of CLP-specific data, makes fine-tuning these large models challenging. To address this, we leverage recent advances in \emph{parameter-efficient fine-tuning} (PEFT)\cite{peft}, which enable effective adaptation to low-resource domains without updating the full model. Specifically, we fine-tune foundation models using \emph{low-rank adaptation (LoRA)}~\cite{lora} and \emph{weight-decomposed low-rank adaptation (DoRA)}~\cite{dora} with a classification objective to enhance CLP detection and severity classification performance.

\section{Parameter Efficient Finetuning}\label{peft}

PEFT refers to a class of techniques that fine-tune large models more efficiently by significantly reducing the number of trainable parameters~\cite{peft}. Instead of updating the entire network, PEFT methods adapt only specific parts of the model—typically the query, key, and value projection matrices within the transformer's attention blocks~\cite{peft2,lora}. By limiting updates to a small subset of parameters, these methods enable effective adaptation in low-resource settings with limited in-domain data~\cite{peft}. In this work, we adopt two PEFT methods: LoRA~\cite{lora} and DoRA~\cite{dora}. We describe each approach briefly in the following subsections.

\subsection{LoRA (Low-Rank Adaptation)}

\begin{figure}
\includegraphics[width=1\linewidth]{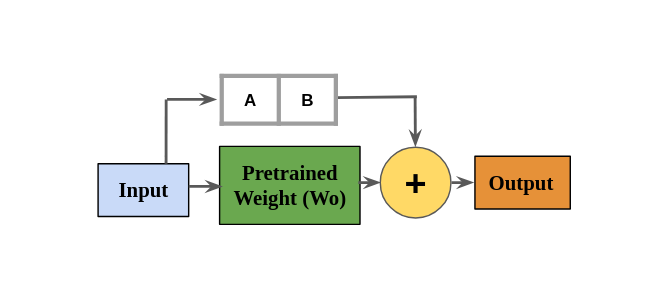}
\caption{Block Diagram of LoRA Framework} \label{fig2}
\end{figure}

LoRA~\cite{lora} is one of the most widely used PEFT methods. It functions as an adapter by introducing trainable parameters only during training, while keeping the model’s original size unchanged. Instead of updating the full weight matrix, LoRA decomposes the weight update 
$\Delta W$ into two smaller low-rank matrices, 
$A$ and $B$, and learns these through backpropagation. After training, it combines them to approximate the full weight update. Given a pre-trained weight matrix $ W_0 \in \mathbb{R}^{d \times k}$, LoRA applies a low-rank decomposition to the weight update such that $\Delta W = BA$, where $A \in \mathbb{R}^{d \times r} $ and $B \in \mathbb{R}^{r \times k}$, with  $r \ll \min(d, k)$. The model then updates the weight matrix as follows:

\begin{equation}
W = W_0 + BA 
\end{equation}

The block diagram of the weight update procedure using LoRA is depicted in Figure~\ref{fig2}.


\subsection{DoRA (Weight-Decomposed Low-Rank Adaptation)}
\begin{figure}
\includegraphics[width=1\linewidth]{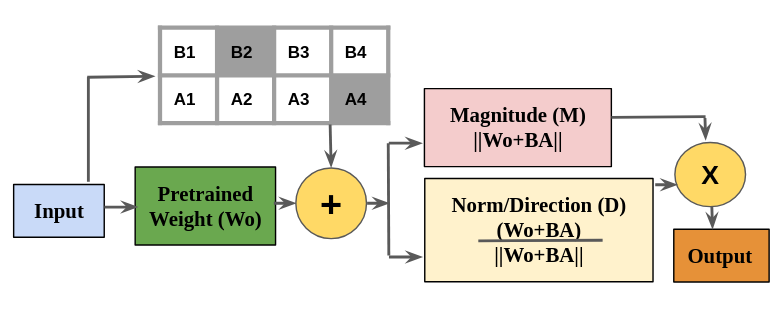}
\caption{Block Diagram of DoRA Framework} \label{fig3}
\end{figure}

DoRA extends LoRA by decomposing the weight update into two distinct components: magnitude and direction. It refines LoRA’s approach by breaking down high-rank LoRA layers into multiple structured single-rank components. During training, DoRA dynamically prunes the less important components, optimizing the parameter budget by retaining only those that contribute meaningfully to the task.Instead of learning a full low-rank update, DoRA uses $r$ pairs of single-rank matrices and continuously evaluates their utility. It removes those with minimal impact, enabling a more efficient and compact adaptation. The updated weight matrix is defined as:
\begin{equation}
W =  M \cdot D,
\end{equation}
where $ M ={\|W_0 + BA \|}$ is a learnable scalar or vector that controls the magnitude of the update, while $D=\frac{W_0 + BA }{\|W_0 + BA \|}$ is the normalized direction of the low-rank update. Figure~\ref{fig3} illustrates the block diagram of the weight update process using DoRA.

\subsection{Proposed PEFT with foundation models}

\begin{figure}
\centering
\includegraphics[height= 190pt,width=245pt]{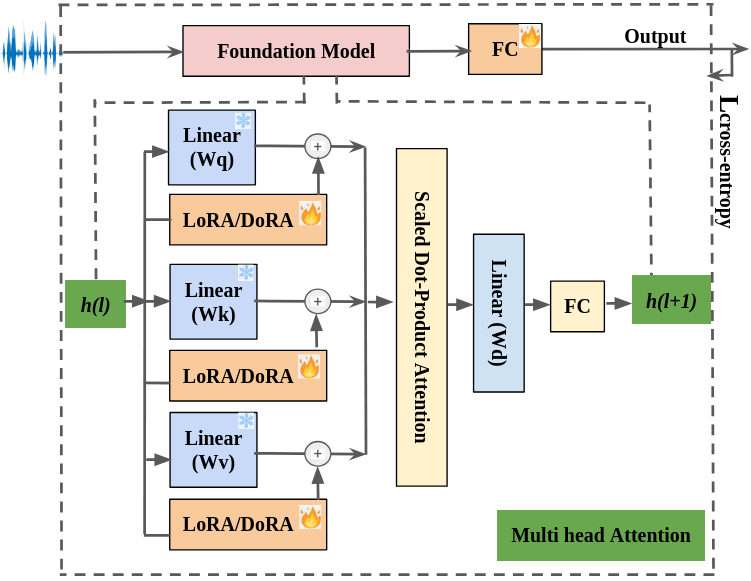}
\caption{Proposed framework: Audio input processed by a foundation model, followed by mean pooling and FC layers to generate logits with cross-entropy loss. LoRA/DoRA updates linear layers (Wq, Wk, Wv) in multi-head attention while freezing other parameters, enhanced by scaled dot-product attention.} \label{fig4}
\end{figure}

We are using the foundation model encoder with either LoRA or DoRA adapters. After the final transformer layer, we add a fully connected layer for classification. We apply mean pooling to the output of the last transformer layer to obtain an utterance-level representation, which we feed into the classification layer. During training, we optimize the LoRA or DoRA adapters using cross-entropy loss, while keeping the rest of the encoder frozen. The block diagram of the proposed training procedure is shown in Figure~\ref{fig4}.



\section{Experimental Setup}\label{experiments}
We conduct our experiments using two datasets: (1) the New Mexico Cleft Palate Centre (NMCPC) dataset~\cite{jawed}, which contains English speech recordings, and (2) the All India Institute of Speech and Hearing (AIISH) dataset~\cite{hypernasality}, which includes Kannada speech recordings. We perform two tasks: \emph{CLP detection} (a binary classification between normal and CLP) and \emph{severity classification} (a four-class classification: normal, mild, moderate, and severe). To establish a baseline, we extract acoustic features using the extended Geneva Minimalistic Acoustic Parameter Set (eGeMAPS)~\cite{egemaps} and Computational Paralinguistics Challenge feature set (ComParE)~\cite{compare} feature sets and train a support vector machine (SVM) classifier~\cite{svm}, inspired by their proven effectiveness in prior biomedical speech studies. Next, we extract pooled representations from the final transformer layer of three foundation models—Whisper\footnote{\label{hfnote}https://huggingface.co/openai/whisper-small}, Wav2Vec2\footnote{https://huggingface.co/facebook/wav2vec2-base-960h}, and WavLM\footnote{https://huggingface.co/microsoft/wavlm-base}—and use these representations to train SVM classifiers. After identifying the best-performing foundation model, we fine-tune it by introducing a classification layer on top and applying parameter-efficient tuning methods using LoRA and DoRA adapters.

\begin{figure}
\centering
\includegraphics[height= 100pt,width=230pt]{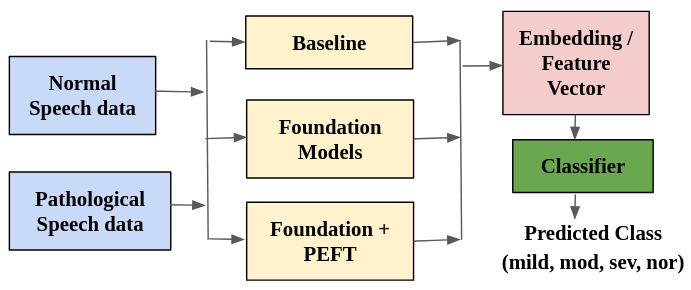}
\caption{Block Diagram of the Proposed Framework} \label{fig5}
\end{figure}

Figure~\ref{fig5} illustrates the overall experimental framework. We evaluate model performance using both accuracy and macro-averaged F1 score, giving special importance to the macro-averaged F1 score due to class imbalance. For all experiments, we pool the training and development sets and apply $5$-fold cross-validation. We select the model that achieves the highest macro-averaged F1 score for final evaluation on the evaluation set.

\subsection{Datasets}

\textbf{NMCPC dataset:} The New Mexico Cleft Palate Centre collected a speech dataset consisting of $65$ speakers, including $41$ speakers with CLP—$22$ male and $19$ female—and $24$ normal speakers—$20$ male and $4$ female. The dataset includes $76$ unique utterances, each containing a maximum of $5$ words, and features speakers aged between $9$ and $13$ years. Clinicians classified the CLP speakers into three severity levels: mild, moderate, and severe~\cite{jawed}. The dataset comprises a total of $1,463$ utterances, which are randomly divided into training, development, and evaluation sets. The training set contains $929$ utterances, including $280$ normal, $246$ mild, $204$ moderate, and $199$ severe. The development set includes $235$ utterances, with $70$ normal, $62$ mild, $53$ moderate, and $50$ severe. The evaluation set comprises $299$ utterances, consisting of $89$ normal, $77$ mild, $67$ moderate, and $66$ severe.



\textbf{AIISH dataset:} The All India Institute of Speech and Hearing, India, collected a speech dataset comprising $60$ speakers, including $31$ normal and $29$ CLP speakers. Among them, $19$ normal female and $12$ normal male speakers participated, along with $9$ CLP female and $20$ CLP male speakers. The dataset includes $19$ unique utterances, each containing a maximum of $3$ words. All participants were native Kannada speakers between the ages of $7$ and $12$ years and did not exhibit any other congenital syndromes such as hearing impairment. The dataset contains a total of $2,726$ utterances and is partitioned into training, development, and evaluation sets. The training set consists of $1,731$ utterances, including $1,106$ normal, $302$ mild, $247$ moderate, and $76$ severe utterances. The development set includes $508$ utterances, comprising $357$ normal, $76$ mild, $56$ moderate, and $19$ severe utterances. The evaluation set contains $487$ utterances, with $278$ normal, $95$ mild, $76$ moderate, and $38$ severe utterances.

\subsection{Baseline Features and Models}
For baseline comparison, we extract two traditional feature sets using openSMILE: eGeMAPS\footnote{{\label{hanote}https://audeering.github.io/opensmile/}}, which provides $88$ dimensional features~\cite{egemaps}, and ComParE\textsuperscript{\ref{hanote}}, which yields $6,373$ dimensional features~\cite{compare}. We also create a merged set that combines both, resulting in a total of $6,461$ dimensional features. To improve computational efficiency, we apply Principal Component Analysis (PCA) on ComParE and the merged set to reduce the handcrafted features to $100$ dimensions. We then use these feature sets with the SVM classifier with  Radial Basis Function (RBF)~\cite{RBF} kernel to evaluate performance.

\subsection{Foundation Models}
We use three state-of-the-art pre-trained models Whisper small~\cite{whispersmall}, Wav2Vec2-base~\cite{wav2vec2}, and WavLM-base~\cite{wavlm} as frozen feature extractors. For each model, we first resample the raw audio to $16$ kHz and then pad or truncate it to a uniform duration of $30$ seconds. For Whisper, we convert the audio into log-Mel spectrograms using its built-in AutoProcessor. We then extract the last hidden states from the encoder and apply mean pooling to obtain a typically $768$-dimensional representation per utterance. We use these pooled representations as input features to an SVM classifier with an RBF kernel for classification.

\subsection{Foundation Models with PEFT}

We use the best-performing foundation model augmented with LoRA and DoRA. We introduce the low-rank adapters only in the key, value, and query matrices of the transformer, using a rank of $8$. We connect the output of the final transformer encoder to a fully connected layer of size $2$ or $4$, depending on whether the task is detection or severity classification. We train the model using the AdaM optimizer~\cite{adam} with a learning rate of $8e{-}5$ and categorical cross-entropy loss.

\section{Results and Discussion}\label{results}
Table~\ref{tab:performance_comparison} presents the performance of both detection and severity classification using the baseline and proposed methods.


\begin{table*}[ht]
\centering
\caption{Performance comparison of models across NMCPC and AIISH datasets for CLP detection (normal and clp) and 4-class severity classification task (mild, moderate, severe, and normal)}
\label{tab:performance_comparison}
\resizebox{\textwidth}{!}{%
\begin{tabular}{|l|l|c|c|c|c|c|c|c|c|}
\hline
\multirow{3}{*}{} & \multirow{3}{*}{\textbf{Features}} & \multicolumn{4}{c|}{\textbf{Detection (2 class)}} & \multicolumn{4}{c|}{\textbf{Severity  classification (4 class)}} \\
\cline{3-10}
& & \multicolumn{2}{c|}{\textbf{NMCPC}} & \multicolumn{2}{c|}{\textbf{AIISH}} & \multicolumn{2}{c|}{\textbf{NMCPC}} & \multicolumn{2}{c|}{\textbf{AIISH}} \\
\cline{3-10}
& & \textbf{Accuracy (\%)} & \textbf{F1 score} & \textbf{Accuracy (\%)} & \textbf{F1 score} & \textbf{Accuracy (\%)} & \textbf{F1 score} & \textbf{Accuracy (\%)} & \textbf{F1 score} \\
\hline
\hline
\multirow{3}{*}{{Baseline Features}} & eGeMAPS & \textbf{81.61} & \textbf{0.80} & \textbf{85.63} & \textbf{0.85} & \textbf{46.82} & \textbf{0.41} & \textbf{60.99} & \textbf{0.34} \\
\cline{2-10}
& ComParE & 56.52 & 0.53 & 66.94 & 0.64 & 27.09 & 0.23 & 51.95 & 0.26 \\
\cline{2-10}
& Merged eGeMAPS \& ComParE & 60.87 & 0.56 & 67.56 & 0.65 & 25.08 & 0.23 & 52.77 & 0.28 \\
\hline
\hline
\multirow{3}{*}{{Foundation models}} & Whisper & \textbf{93.31} & \textbf{0.92} & \textbf{93.02} & \textbf{0.93} & \textbf{56.86} & \textbf{0.53} & \textbf{70.02} & \textbf{0.49} \\
\cline{2-10}
& Wav2vec2 & 81.61 & 0.79 & 87.89 & 0.87 & 43.81 & 0.40 & 67.97 & 0.46 \\
\cline{2-10}
& WavLM & 78.26 & 0.71 & 92.20 & 0.92 & 41.81 & 0.40 & 64.27 & 0.36 \\
\hline\hline
\multirow{2}{*}{{\begin{tabular}{l}\textbf{Foundation model+}\\\textbf{LoRA/DoRA}\end{tabular}}} 
& Whisper LoRA & 93.65 & 0.93 & \textbf{94.25} & \textbf{0.94} & 68.23 & 0.65 & \textbf{72.28} & \textbf{0.52} \\
\cline{2-10}
& Whisper DoRA & \textbf{94.31} & \textbf{0.93} & 93.63 & 0.93 & \textbf{70.23} & \textbf{0.67} & 69.82 & 0.51 \\
\hline
\end{tabular}%
}
\end{table*}


\subsection{CLP Detection}
Table~\ref{tab:performance_comparison} reveals that the eGeMAPS feature set outperforms both ComParE and the merged feature set on both datasets. On the NMCPC dataset, eGeMAPS achieves a classification accuracy of $81.61\%$ and an F1 score of $0.80$. In comparison, ComParE achieves only $56.52\%$ accuracy and an F1 score of $0.53$, while the merged feature set yields $60.87\%$ accuracy and an F1 score of $0.56$. On the AIISH dataset, eGeMAPS again delivers the best performance, achieving $85.63\%$ accuracy and a $0.85$ F1 score. ComParE performs lower, with $66.94\%$ accuracy and a $0.64$ F1 score, and the merged set slightly improves to $67.56\%$ accuracy and a $0.65$ F1 score. These results clearly show that eGeMAPS provides better discrimination between normal and CLP speech than ComParE and its combination with eGeMAPS.

We also observe improved performance with foundation models, where Whisper outperforms both Wav2Vec2 and WavLM. On the NMCPC dataset, Whisper achieves $93.31\%$ accuracy and an F1 score of $0.92$, while Wav2Vec2 attains $81.61\%$ accuracy and $0.79$ F1 score, and WavLM yields $78.26\%$ accuracy with an F1 score of $0.71$. We notice a similar trend on the AIISH dataset. Whisper achieves $93.02\%$ accuracy and a $0.93$ F1 score, outperforming Wav2Vec2 with $87.89\%$ accuracy and a $0.87$ F1 score, and WavLM with $92.2\%$ accuracy and a $0.92$ F1 score. These results indicate that among the foundation models, Whisper provides superior discriminative ability in classifying CLP and normal speech.

Motivated by Whisper’s promising performance among foundation models, we evaluated the proposed LoRA and DoRA adaptation methods within the Whisper model. On the NMCPC dataset, Whisper with DoRA achieved a higher accuracy of $94.31\%$ and an F1 score of $0.93$, compared to $93.65\%$ accuracy and the same F1 score of $0.93$ using LoRA. However, on the AIISH dataset, LoRA outperformed DoRA, achieving $94.25\%$ accuracy and a $0.94$ F1 score, while DoRA achieved $93.63\%$ accuracy and a $0.93$ F1 score. This reversal in performance between datasets may be attributed to the difference in language—NMCPC contains English speech, while AIISH contains Kannada. Further analysis is necessary in this direction to draw a more detailed and conclusive explanation.

\subsection{Severity Classification}

We observe a similar trend in severity classification as in the detection task. However, the performance of severity classification is comparatively lower, likely due to the increased complexity of distinguishing between four classes instead of two. Among the baseline feature sets, eGeMAPS achieves better results, with an F1 score of $0.41$ on the NMCPC dataset and $0.34$ on the AIISH dataset. In contrast, ComParE yields lower F1 scores of $0.23$ and $0.26$ on NMCPC and AIISH, respectively, while the merged ComParE and eGeMAPS features achieve $0.23$ and $0.28$. These results indicate that, similar to the detection task, eGeMAPS offers better discriminative ability for differentiating between severity levels compared to ComParE and the combined feature set.

Using the foundation models, we observe that Whisper embeddings combined with an SVM classifier achieve the best performance in severity classification, similar to the detection task. Whisper attains F1 scores of $0.53$ on the NMCPC dataset and $0.49$ on the AIISH dataset. In comparison, Wav2Vec2 achieves F1 scores of $0.40$ and $0.46$, while WavLM yields $0.40$ and $0.36$ on NMCPC and AIISH, respectively. These results indicate that Whisper embeddings offer superior discriminative ability in differentiating between severity levels compared to Wav2Vec2 and WavLM.

Finally, severity classification follows a similar trend to CLP detection using proposed parameter-efficient fine-tuning with Whisper. On the NMCPC dataset, DoRA achieves a slightly better F1 score of $0.67$ compared to $0.65$ with LoRA. In contrast, on the AIISH dataset, LoRA performs slightly better, achieving an F1 score of $0.52$, while DoRA yields $0.51$. These results further support the observation that adaptation methods may behave differently across datasets, possibly due to language variations.

\subsection{Discussions}
We began our experiments with baseline features for CLP detection and severity classification. Among these, eGeMAPS combined with an SVM classifier achieved the best performance in terms of both accuracy and F1 score on both the NMCPC and AIISH datasets. We then evaluated foundation models, including the self-supervised Wav2Vec2 and WavLM, as well as the weakly supervised Whisper model. For both detection and severity classification, Whisper embeddings consistently outperformed the others in terms of accuracy and F1 score. Motivated by Whisper’s strong performance, we applied our proposed parameter-efficient fine-tuning methods using LoRA and DoRA. We observed that DoRA yielded better results on the NMCPC dataset, while LoRA performed better on the AIISH dataset.

\begin{figure}
\centering
\includegraphics[height= 190pt,width=250pt]{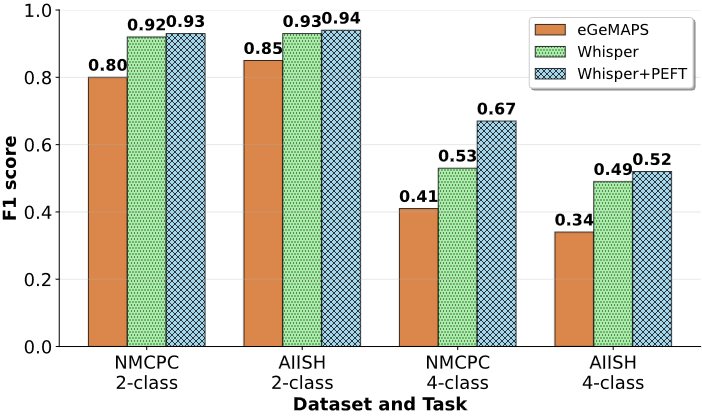}
\caption{Performance comparison (F1-scores) of the best-performing models using eGeMAPS, Whisper, and Whisper+PEFT feature sets across the NMCPC and AIISH datasets for CLP detection and severity classification tasks. For AIISH (both tasks), the Whisper+PEFT model uses LoRA, while for NMCPC (both tasks), the Whisper+PEFT model uses DoRA.} \label{fig6}
\end{figure}

We compared the performance across methods and found that our proposed LoRA and DoRA adaptations applied to the Whisper model outperformed both the pretrained Whisper with SVM and the traditional eGeMAPS features with SVM in CLP detection and severity classification on both the NMCPC and AIISH datasets. On the NMCPC dataset, Whisper with DoRA achieved an F1 score of $0.93$ for CLP detection and $0.67$ for severity classification. In comparison, the pretrained Whisper with SVM achieved F1 scores of $0.92$ and $0.49$, while eGeMAPS with SVM reached $0.80$ and $0.41$. On the AIISH dataset, Whisper with LoRA attained F1 scores of $0.94$ for CLP detection and $0.52$ for severity classification, compared to $0.93$ and $0.49$ from Whisper with SVM, and $0.85$ and $0.34$ from eGeMAPS with SVM. Figure~\ref{fig6} presents a visual comparison of these results. These findings confirm our initial hypothesis that applying LoRA and DoRA for parameter-efficient fine-tuning improves performance in both CLP detection and severity classification over the use of pretrained foundation models and handcrafted feature-based approaches.

\balance
\section{Conclusions}\label{conclusions}

We conducted a comprehensive study on CLP detection and severity classification using both traditional acoustic features and representations from foundation models. Our experiments on the NMCPC and AIISH datasets showed that eGeMAPS features outperformed ComParE when used with SVM classifiers. Among the foundation models, Whisper consistently achieved higher accuracy and F1 scores than Wav2Vec2 and WavLM. Motivated by Whisper’s strong performance, we applied parameter-efficient fine-tuning using LoRA and DoRA adapters. Whisper with DoRA yielded the best results on the English NMCPC dataset, while LoRA performed better on the Kannada AIISH dataset.

These findings highlight the effectiveness of the proposed parameter-efficient fine-tuning strategy for foundation models in both CLP detection and severity classification. In future work, we plan to investigate language-specific fine-tuning to better understand the impact of language characteristics on model performance. We also aim to explore explainability techniques to interpret the model’s decisions, particularly in distinguishing between CLP severity levels. These directions will help us develop more transparent, inclusive, and clinically useful models for speech-based health diagnostics.



\bibliographystyle{plain} 


\end{document}